%% file: main.tex
\documentclass[11pt]{article}

\usepackage[margin=1in]{geometry}
\usepackage{amsmath, amssymb, amsfonts}
\usepackage{graphicx}
\usepackage{algorithm}
\usepackage{algorithmic}
\usepackage{subcaption}

\title{Implicitly Parallel Neuromorphic Solver Design for Constraint Satisfaction Problems}
\author{
  Recep Bugra Uludag \\
  University of Minnesota - Twin Cities \\
  \texttt{uluda002@umn.edu}
  \and
  Ahmet Efe \\
  University of Minnesota - Twin Cities \\
  \texttt{efe00002@umn.edu}
  \and
  Ismail Akturk \\
  Ozyegin University \\
  \texttt{ismail.akturk@ozyegin.edu.tr}
  \and
  Ulya R. Karpuzcu \\
  University of Minnesota - Twin Cities \\
  \texttt{ukarpuzc@umn.edu}
}

\date{}

\begin{document}
\maketitle

\begin{abstract}
Many real-life problems of practical importance -- spanning a wide range of applications from chip design to bioinformatics --  represent constraint satisfaction problems, where classical solvers have to rely on heuristic approximations due to the computational complexity. Neuromorphic solvers, on the other hand, offer a unique alternative representation which enables an inherently parallel 
exploration of the solution space. This paper provides a theoretical characterization and experimental demonstration of this native type of parallelism that is hard to apply to classical solvers. We observe that more than two orders of magnitude faster operation is possible without compromising solution accuracy. Our study represents the first step toward bridging the theory vs. practice gap to unlock the performance potential of emerging neuromorphic solvers. 
\end{abstract}

\section{Introduction}\label{Introduction}
\input{Texts/Introduction}

\section{Background} \label{Background}
\input{Texts/Background.tex}
\section{Toward an Inherently Parallel Neuromorphic Solver for CSPs} \label{Methodology}
\input{Texts/Methodology.tex}

\section{Evaluation Setup} \label{Evaluation Setup}
\input{Texts/Evaluation_Setup.tex}

\section{Evaluation} \label{Evaluation}
\input{Texts/Evaluation.tex}

\section{Related Work} \label{Related Work}
\input{Texts/Related_Work.tex}

\section{Conclusion} \label{Conclusion}
\input{Texts/Conclusion.tex}

\bibliographystyle{ieeetr}
\bibliography{references}

\end{document}

%% file: Texts/Introduction.tex
\noindent Constraint Satisfaction Problems (CSPs) form a unifying framework for a wide range of computational problems. At their core, CSPs abstract problems into a set of variables, domains, and constraints, providing a standard representation that transcends application boundaries. This universality makes CSPs central to both theoretical computer science and practical problem-solving: many computational tasks -- such as formal verification in VLSI design, resource allocation in distributed systems, and automated planning in artificial intelligence -- can be systematically formulated as CSPs. Beyond their modeling power, CSPs are computationally challenging -- many formulations are NP-hard or NP-complete.

While classical CSP solvers on CPUs have achieved remarkable success, their energy efficiency remains limited.
This motivates the exploration of emerging computing paradigms tailored to the computational needs of CSPs such as Neuromorphic Computing (NC). NC systems draw inspiration from the brain, leveraging massively parallel, event-driven architectures that integrate memory and processing in the same physical substrate \cite{Hamdioui_Data_Intensive_2015,Pedretti_phase_change_2020}. Inherent near-memory or in-memory computation reduces data movement and along with the inherent parallelism, opens the door to practically efficient CSP solvers at scale\cite{Pedretti_phase_change_2020,Aimone_non_cognitive_NC_2022,Davies_Advancing_2021}.

Classical solvers rely not only on systematic search but also on an extensive repertoire of heuristics -- such as 
variable/value-ordering \cite{Elliot_Tree_Search_1980}, conflict-driven learning \cite{Marques_Clause_Conflict_1999}, and local consistency checks \cite{Mackworth_Consistency_1977}
-- to detect and leverage common structural patterns in CSPs. Such heuristics typically result in dramatic improvements in solver performance.
In contrast, practical neuromorphic CSP solvers are still in their early stages where most of the efforts go to neuromorphic formulations of specific CSPs and proof-of-concept demonstrations. This paper aims to bridge the theory vs. practice gap towards unlocking the performance potential of neuromorphic solvers at scale by introducing novel heuristics leveraging the inherent parallelism in NC. 
Specifically, based on unique representational properties of NC where multiple stable states to encode CSP solutions can coexist, our approach allows the solver to explore multiple feasible solutions at a time, \textit{in parallel} -- a capability difficult to realize in traditional architectures. 

In the following, we provide a formal characterization of the proposed heuristic as well as a simulation-based evaluation considering a representative subset of CSPs.

%% file: Texts/Background.tex
\subsection{Constraint Satisfaction Problems (CSPs)}

\noindent A Constraint Satisfaction Problem (CSP) is formally defined as a triplet 
$\langle X, D, C \rangle$ \cite{tsang_foundations_1993}, where:
\begin{itemize}
    \item $X = \{X_1, \dots, X_N\}$ is a finite set of variables,
    \item $D = \{D_1, \dots, D_N\}$ is a set of finite domains, with each $X_i$ taking values from $D_i$,
    \item $C = \{C_1, \dots, C_m\}$ is a set of constraints. Each constraint 
    $C_j$ is a relation defined over a subset of the variables, specifying the 
    allowed combinations of values for those variables.
\end{itemize}
A domain is the set of permissible values for a variable, e.g., 
$\{0,1\}$ for Boolean satisfiability (SAT) or $\{r,g,b\}$ (\textit{red}, \textit{green}, and \textit{blue}, respectively) in graph (3-)coloring problems. An \textit{assignment} is a mapping from a subset of variables to values in their domains. A \textit{complete assignment} assigns a value to every variable in $X$. 
A solution to a CSP is a complete assignment such that all constraints in $C$ are simultaneously satisfied.
Graphs commonly serve visual representation of CSPs: In a standard constraint graph, each node corresponds to a variable; and each edge, to a constraint between the respective nodes, i.e., variables. 

As representative examples, Boolean Satisfiability (SAT) and graph coloring are common CSPs with numerous practical applications:
\begin{list}{\labelitemi}{\leftmargin=1em}  
    \item SAT: Given Boolean variables $X=\{x_1,\dots,x_N\}$ with domains $\{0,1\}$ and a formula $F=\bigwedge_{j=1}^m C_j$ (termed CNF, Conjunctive Normal Form) where each clause $C_j$ is a disjunction of literals (i.e., Boolean variable $x_i$ or its negation $ \lnot x_i$), decide whether there exists an assignment $a:X\to\{0,1\}$ such that $F(a)=\text{true}$ (equivalently, all clauses are satisfied).
    \item Graph $k$-coloring: Given a graph $G=(V,E)$, define one variable $X_v$ for each vertex $v\in V$ with domain $\{1,\dots,k\}$ (where each value corresponds to a color). For every edge $(u,v)\in E$ impose the binary constraint $X_u\neq X_v$. A solution is a proper $k$-coloring of $G$.
\end{list}

Classical CSP solvers typically use \textit{systematic search}, where heuristic strategies to prune the search space -- such as backtracking combined with constraint propagation {\cite{Sabin_Constraint_Propagation_Backtracking_1994}} -- are necessary as the search space exponentially grows with the problem size. 
\textit{Local search}, on the other hand, begins with a complete assignment and iteratively improves it by modifying the assignment --  by changing the values for a limited number of variables, guided by approximation heuristics. While incomplete -- unable to guarantee a solution or prove that no solution exists --
local search methods often outperform systematic search approaches for large-scale CSPs. Variants range from deterministic hill-climbing {\cite{Selman_GSAT_1992}} to stochastic methods such as {random walks and simulated annealing} {\cite{selman_noise_1994}} and their efficiency in exploring large regions of variables makes them particularly attractive at scale.

\subsection{Neural Sampling} 

\noindent 
Most of the Neuromorphic CSP solvers fundamentally rely on \textit{Neural Sampling}, that enables probabilistic inference by Markov Chain Monte Carlo (MCMC) sampling in Spiking Neural Networks (SNNs) \cite{buesing_neural_2011}. Its theoretical foundation is closely related to Boltzmann Machines \cite{Boltzmann_1989}, but extends this framework to more
realistic spiking neuron dynamics.

A Neural Sampling network consists of \textit{principal} neurons 
$x = \{x_1,\dots,x_N\}$, where each neuron is 
a binary random variable corresponding to one value in the domain of a CSP variable. For clarity, we will refer to these as the variable’s \textit{domain} neurons. The state of a neuron is defined over a short time window of length $\tau$ (a fixed duration parameter defining how long a spike is registered): if neuron $i$ has emitted a spike within the interval $(t-\tau,t]$, then $x_i(t)=1$, otherwise $x_i(t)=0$. The instantaneous network state is therefore the binary vector $x(t) = (x_1(t),\dots,x_N(t)) \in \{0,1\}^N$.

The membrane potential of {neuron $i$} is modeled as
\begin{equation}
    u_i(t) = b_i + \sum_j w_{ij} x_j(t),
\end{equation}
where $b_i$ is a bias and $w_{ij}$ is the synaptic weight from neuron $j$.
Stochastic spiking is introduced by defining the firing probability as
\begin{equation}
    P(x_i(t)=1 \mid x_{\backslash i}(t)) = \frac{1}{\tau} e^{u_i(t)},
\end{equation}
which satisfies the Neural Computability Condition (NCC)
\cite{buesing_neural_2011}. The NCC requires that a neuron’s
membrane potential encodes the log-odds of its variable being active, i.e.,

\begin{equation}
    u_i(t) = \log \frac{p(x_i(t)=1 \mid x_{\backslash i}(t))}{p(x_i(t)=0 \mid x_{\backslash i}(t))}
\end{equation}
ensuring that the network dynamics implement the correct sampling process.
$x_{\backslash i}(t)$ denotes the instantaneous values 
$x_j(t)$ of all other variables with $j \neq i$.
The stochasticity is crucial as a computational element.

The asynchronous network dynamics induce a Markov chain over states $x(t)$.
Although spike-based transitions are not reversible (a neuron enters state $1$ stochastically but returns to $0$ deterministically after $\tau$), the chain is irreducible (i.e., any state can be reached from any other state in a finite number of transitions) and aperiodic, and convergence to a unique stationary distribution is guaranteed as long as the NCC is satisfied
\cite{buesing_neural_2011}.
The resulting stationary distribution is a Boltzmann distribution of the form
\begin{equation}
    p(x(t)) \propto \exp\!\Bigg(\sum_i b_i x_i(t) +
\tfrac{1}{2}\sum_{i,j} w_{ij} x_i(t) x_j(t)\Bigg).
\end{equation}
Equivalently, this can be expressed as
\begin{align}
    p(x(t)) &\propto e^{-E(x(t))}, \\
    E(x(t)) &= -\sum_i b_i x_i(t)
              - \tfrac{1}{2}\sum_{i,j} w_{ij}\, x_i(t) x_j(t).
\end{align}
so that the network dynamics implement stochastic sampling over an energy landscape $E$ defined by biases $b_i$ and synaptic couplings $w_{ij}$.

This probabilistic sampling provides a natural foundation for solving CSPs: constraints are encoded as additive penalty terms in the network energy $E(x(t))$, so that violated constraints raise the energy -- and thereby reduce the probability of the corresponding assignments being sampled per $p(x(t)) \propto e^{-E(x(t))}$ -- where valid assignments correspond to lower-energy states. 

%% file: Texts/Methodology.tex
\noindent 
{\bf Core Building Blocks -- Neural Motifs:} To illustrate neuromorphic representations of CSPs, we will first introduce two standard neural \textit{motifs} -- small recurring connection blocks that add energy penalties based on the joint state of their connected neurons.

The first is the \textit{Winner-Take-All (WTA)} motif. It consists of a group of neurons, each representing a different option (e.g., for the values of a CSP variable or for mutually exclusive assignments such as in graph coloring). The motif enforces that at most one neuron in the group can be active at a time: when one turns on, it suppresses the others through feedback connections from an auxiliary WTA neuron. Each WTA motif is parameterized by a coupling strength, the \textit{WTA weight}, which sets how strongly the neurons suppress each other. This suppression of a neuron from firing is referred to as \textit{inhibition}. A high weight enforces strict \textit{one-hot} activity, while a low weight allows multiple neurons to remain active temporarily. In energy terms, states with multiple active neurons are penalized, so that only one-hot assignments remain low-energy. For example, in a (non-trivial) SAT problem, two separate neurons represent the two distinct values for each variable -- i.e., one neuron for \textit{True}, another one for \textit{False}. The WTA motif in this case ensures that the energy is low only when exactly one of these neurons is active, as the respective variable cannot typically be both \textit{True} and \textit{False} in a non-trivial problem instance.

\begin{figure*}[htbp]
    \centering
    \includegraphics[width=\linewidth]{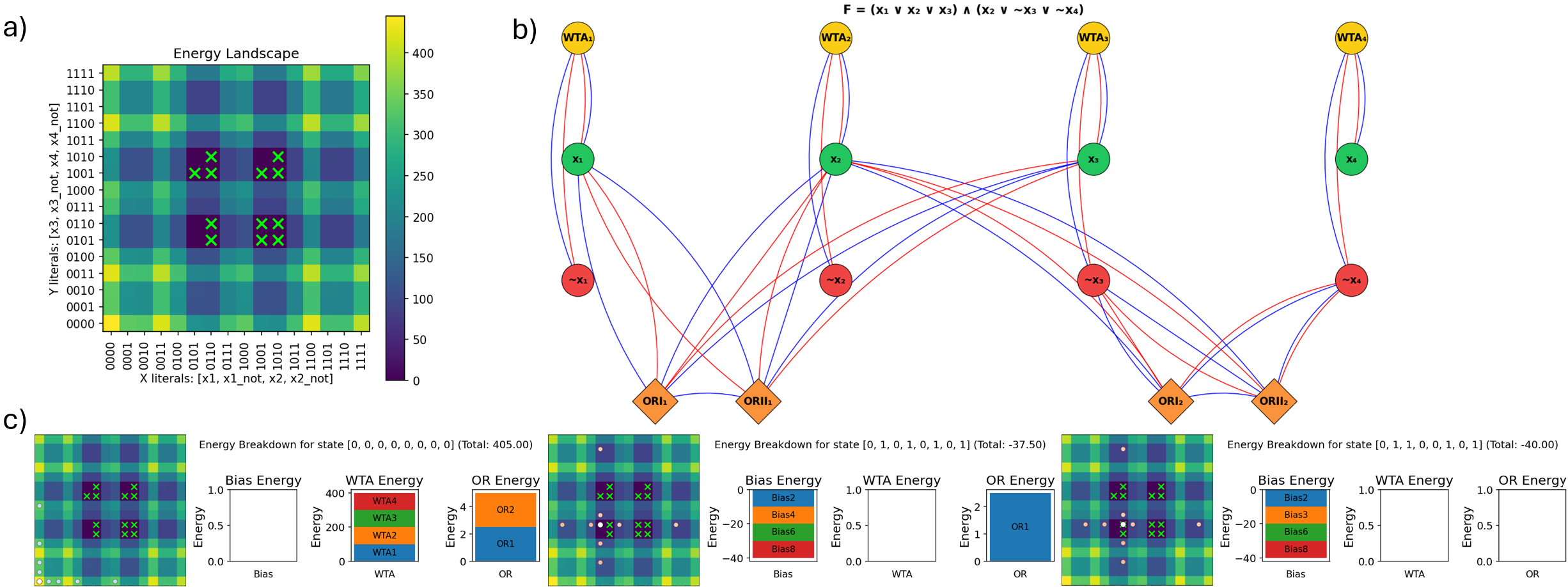}
    \caption{(a) Analytical energy landscape with valid solutions indicated.
             (b) Mapping of a 4-variable, 2-clause SAT instance onto a spiking network using WTA and OR motifs. Blue connections denote positive weights, while red connections denote negative weights.
             (c) Energy breakdown across representative states: a configuration with all constraints violated (highest energy), a local minimum, and a satisfying assignment (lowest energy). The white dot marks the selected state, while the surrounding dots represent its one-bit-Hamming-distance neighbors. Energy differences between the selected state and its neighbors are color-coded: decreases (favorable moves) appear in blue, whereas increases (unfavorable moves) appear in red.}
    \label{fig:Mapping}
\end{figure*}

The second is the \textit{OR} motif. Continuing with the SAT example, this motif enforces clause satisfaction: Applied to a set of neurons representing the literals of a clause, it raises the energy only when all neurons are inactive; if at least one is active, the clause is satisfied and the energy remains low \cite{jonke_solving_2016}. In practice, this is realized with a pair of auxiliary neurons (such as the ones labeled as ORI/ORII in Fig.\ref{fig:Mapping}) whose joint effect pushes neurons that represent the literals to fire when the clause is unsatisfied.

\noindent {\bf A Representative Example:} Fig.\ref{fig:Mapping} (panel b) illustrates how the SAT instance \\
\indent  \indent  \indent$(x_1 \lor x_2 \lor x_3) \land (x_2 \lor \lnot x_3 \lor \lnot x_4)$\\
\noindent is mapped onto a spiking network. First each variable $x_i$ is represented by a pair of principal neurons encoding $\{\text{True},\text{False}\}$, and each such pair is connected by a WTA motif ($WTA_1-WTA_4$) to enforce exclusivity. The two OR motifs ($OR_1-OR_2$) connect the corresponding literal neurons to encode the two clauses. In addition, bias terms ($Bias_1-Bias_4$) exist to encourage neuron activity in the network.  

The analytical energy landscape (panel a) shows high-energy points where all constraints are violated, local minima for partial solutions, and global minima for satisfying assignments (marked with $\times$), with darker colors associating with lower energy states.  Panel c decomposes the total energy of an indicated state into contributions from bias, WTA, and OR motifs, illustrating how energy is formalized. Through stochastic sampling, the network visits low-energy states more frequently and thereby converges toward valid assignments representing solutions.

\noindent {\bf Parallel Solution Discovery:} Existing neuromorphic CSP solvers based on neural sampling \cite{habenschuss_stochastic_2013, jonke_solving_2016} predominantly represent proof-of-concept designs and typically do not leverage problem structure, nor the native parallelism enabled by neuromorphic representations to enhance solver efficiency -- in contrast to our solver, which we introduce next.

A key distinction between classical and neuromorphic approaches to CSP solving lies in variable representation. In classical solvers, a variable with domain $D$ is represented by a single symbol selected from $D$.
In neuromorphic solvers, however, each domain element is assigned its own {principal (or domain) neuron},
allowing a variable to occupy not only singleton values but also subsets of its domain. These multi-valued assignments enable the network to hold several candidate values simultaneously. When interpreted at the problem level, this allows the solver to explore multiple solutions in parallel, rather than enumerating them sequentially as in classical search.

We leverage this behavior through a structure-aware heuristic, which we will refer to as \textit{parallel discovery heuristic}.
Specifically, we modulate the WTA inhibition over a variable’s domain neurons according to the variable’s connectivity in the constraint graph (where each variable is represented by a node; and each constraint by an edge).
Intuitively, tightly constrained variables (as characterized by a high edge degree in the constraint graph) should commit quickly to single values, while loosely constrained variables (as characterized by a low edge degree in the constraint graph) should be allowed to remain multi-valued, thereby supporting parallel solution discovery.

The applicability of this heuristic is determined by the constraint coverage of each variable. We define a mutual-exclusion constraint as a pairwise exclusion: two domain assignments cannot be active simultaneously -- e.g., adjacent vertices in graph coloring cannot share the same color, enforced by a WTA motif between the corresponding neurons. 
If every domain neuron of a variable is coupled through mutual-exclusion constraints to all other neurons (i.e., forming a $D$-clique), then the variable is forced to a singleton assignment. We formalize this as:

\noindent \textit{Principle-1: In a $D$-clique of mutual-exclusion constraints, each variable is restricted to singleton assignments; no multi-valued assignment is feasible. }

Conversely, if a variable has degree $d < D-1$ in the constraint graph, at least $D-d$ domain values remain unconstrained by neighbors. This guarantees that the variable can safely take a multi-valued assignment.

\noindent \textit{Principle-2: If a variable of domain size $D$ has degree $d < D-1$, then there exists at least one non-singleton assignment disjoint from all neighbors.}

Our heuristic is effective precisely when variables are not fully constrained. For example, graph coloring instances with sparse connectivity admit multi-valued states, whereas problems like Sudoku or dense spin-glass models ({where many variables participate in clique structures}) force singletons.

Accordingly, we define the WTA weight $w_i$ for variable $i$ with degree $n_i$ as:

\begin{equation} \label{eq:Heuristic}
    w_i =
    \begin{cases}
        w_{\max}, & d_i \geq D-1 \\
        \dfrac{d_i}{D-1} \cdot w_{\max}, & 0 \leq d_i < D-1
    \end{cases}
\end{equation}

High-degree variables receive strong inhibition to enforce unique values, while low-degree variables are given weaker inhibition, promoting multi-valued assignments.
In this way, the network dynamically balances constraint satisfaction with exploratory diversity.

\begin{figure}[htbp]
    \centering
    \includegraphics[width=\linewidth]{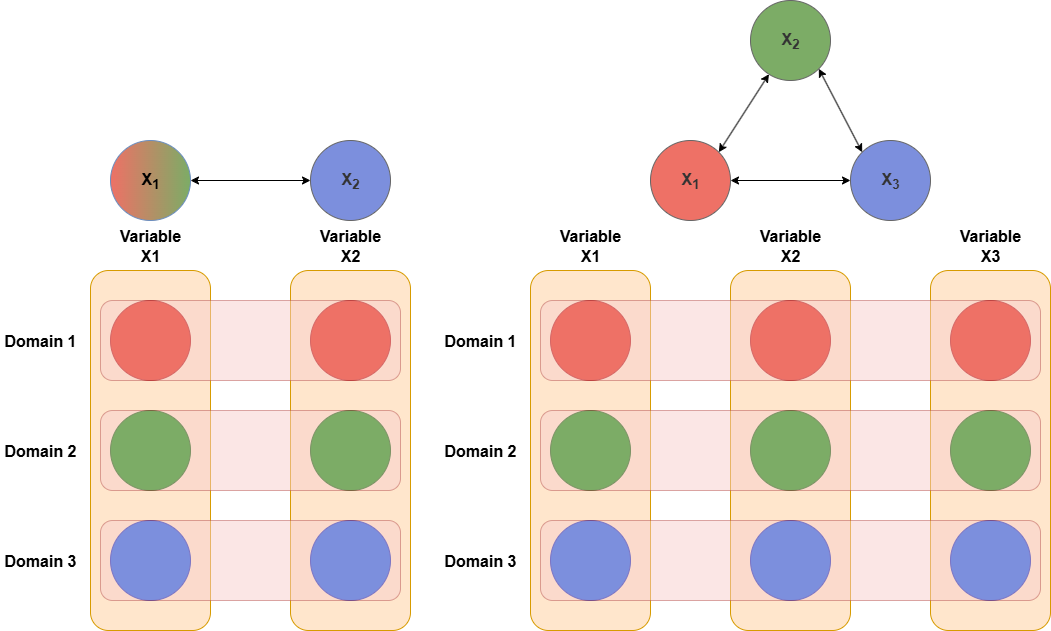}
    \caption{Illustration of two different constraint connectivity in graph coloring with D=3. Each vertical orange box groups the domain neurons (red, green, blue) of a variable, coupled by a WTA motif to enforce one-hot selection. Horizontal connections denote WTA motifs implementing mutual-exclusion constraints between variables. Left: non-clique connectivity allows a variable to remain multi-valued. Right: a full 3-clique in the constraint graph forces singleton assignments. 
    }
    \label{fig:clique_coverage}
\end{figure}

Fig.\ref{fig:clique_coverage} illustrates this principle in a graph coloring setting. With $D=3$, a partially constrained graph allows a variable to adopt multiple colors simultaneously (left), whereas a fully connected clique structure in the constraint graph enforces singleton assignments (right).

Our proposed heuristic thereby adapts SNN dynamics to the structural properties of the CSP graph. It enables parallel exploration of multiple solutions when applicable, while reducing to standard single-solution dynamics when not. Importantly, the heuristic introduces no overhead in inapplicable cases, as high-degree variables naturally collapse to singleton assignments under (\ref{eq:Heuristic}). This property ensures robustness while opening a path toward more efficient solution discovery in under-constrained CSPs.

%% file: Texts/Evaluation_Setup.tex
\noindent 
We evaluate our solver in two stages: 
(i) quantitative characterization of the proposed parallel discovery heuristic;
(ii) comparison to representative neuromorphic CSP solvers. 
All experiments were conducted in 64-bit floating-point precision using Python 3.10.7. The solver is a custom simulator relying on standard scientific libraries (NumPy, Pandas), with classical solvers used solely to verify correctness.

For (i), for a proof-of-concept quantitative demonstration of our parallel discovery heuristic without loss of generality, we use 1000 planar graph coloring problems of 0.8 density with varying node sizes (9, 25, 36, 49). {We adapt problem sizes used in the most related work to facilitate direct apples-to-apples comparison}. Beyond its theoretical importance, graph coloring has numerous practical applications (e.g., register allocation and scheduling), making it a representative test case for evaluating multiple-solution discovery.
We run each problem instance for 100 trials with the heuristic enabled and disabled to isolate the impact of the heuristic on performance, efficiency, robustness, and 
applicability across different problem structures.
For (ii), we consider Jonke et al. \cite{jonke_solving_2016}, Fonseca Guerra et al. \cite{fonseca_guerra_using_2017}, and Rutishauser et al. \cite{rutishauser_solving_2018} using the following benchmark problems:
    {SAT} covering SATLIB uf20-91, uf50-218, uf75-325 \cite{hoos_satlib_2000} 
    {-- random satisfiable instances from the transition region where the hardest satisfiable problems reside};
    {Graph Coloring} considering World (4 colors), Canada (3 colors), Australia (3 colors) maps \cite{fonseca_guerra_using_2017,rutishauser_solving_2018}; 
    {Sudoku} spanning easy, hard, AI Escargot sudoku instances \cite{fonseca_guerra_using_2017}; and
    {Ising} with 10-spin anti-ferromagnetic ring, $10^3$ ferromagnetic cube, and $10^3$ anti-ferromagnetic cube \cite{fonseca_guerra_using_2017}.

%% file: Texts/Evaluation.tex
\subsection{Quantitative Analysis of the Parallel Discovery Heuristic}

\begin{figure}[htbp]
    \centering

    \begin{subfigure}[t]{0.48\linewidth}
        \centering
        \includegraphics[width=\linewidth]{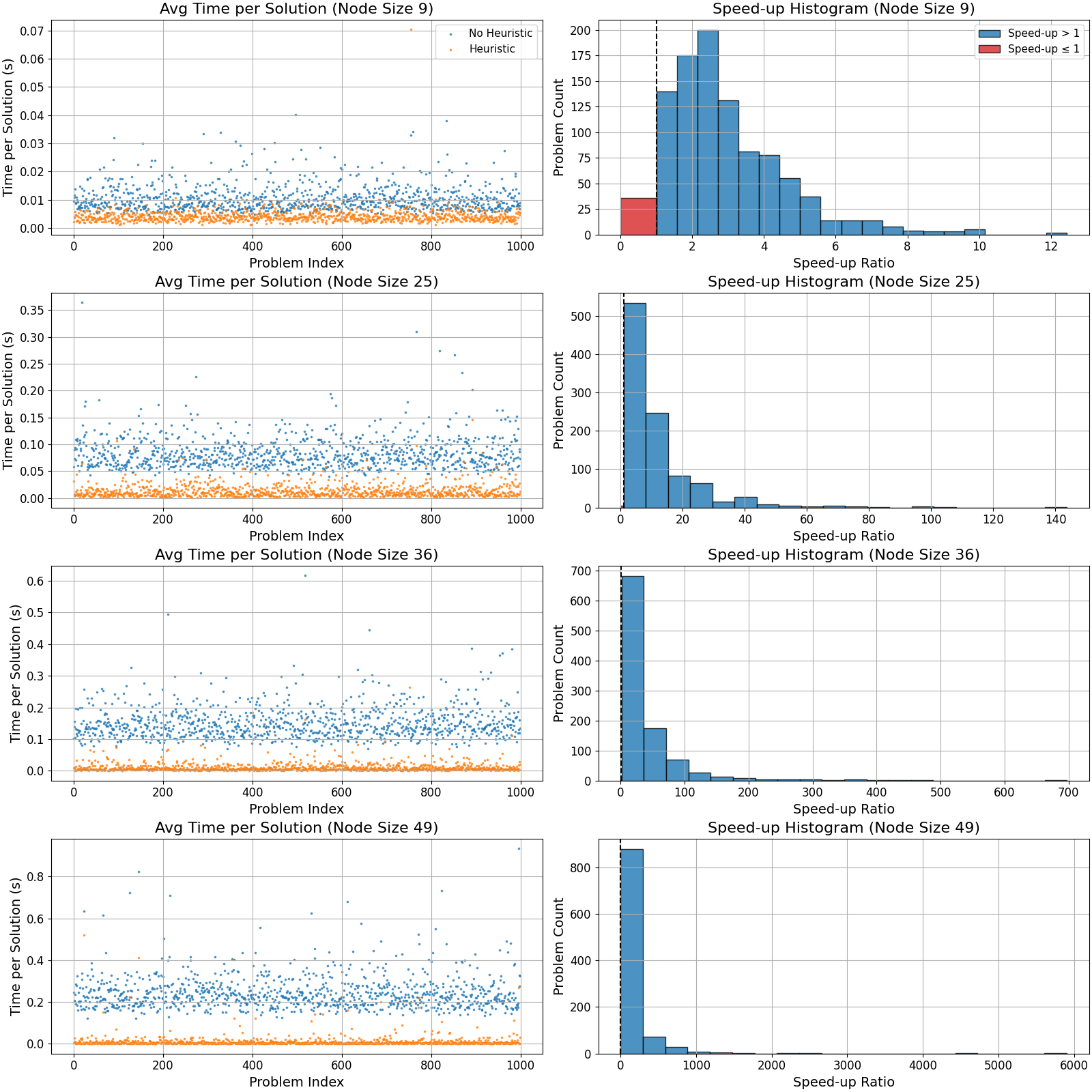}
        \caption{Mean/variance of time per solution (left) and histogram of per-solution speedup (right). Larger, less structured problems feature more degrees of freedom, leading to higher speed-ups.}
        \label{fig:Heuristic_time_per_solution}
    \end{subfigure}
    \hfill

    \begin{subfigure}[t]{0.48\linewidth}
        \centering
        \includegraphics[width=\linewidth]{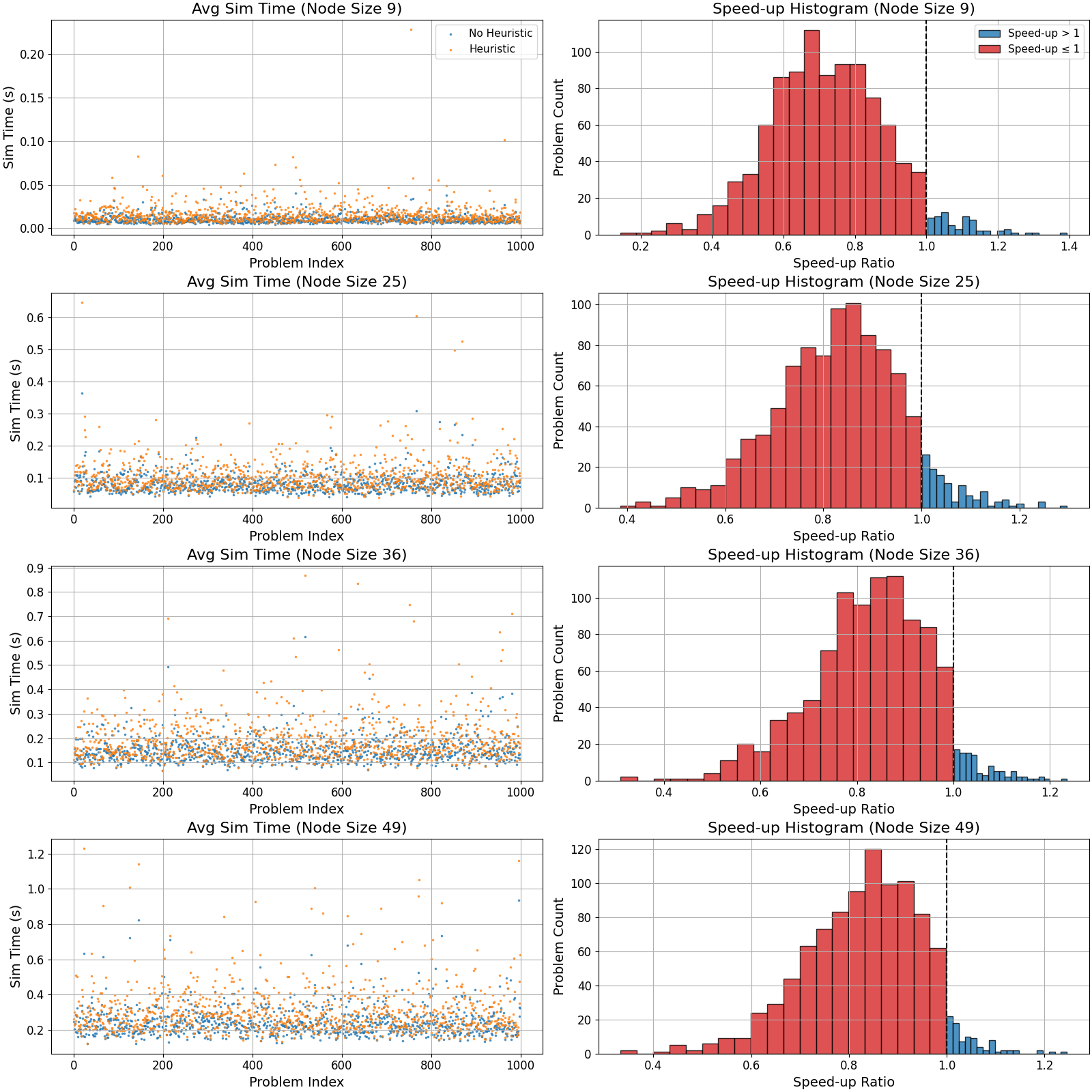}
        \caption{Total time to first solution with and without the heuristic.}
        \label{fig:Heuristic_total_time}
    \end{subfigure}

    \caption{Performance comparison of the proposed heuristic across planar 4-coloring benchmarks.}
    \label{fig:Heuristic_combined}
\end{figure}

\noindent{\bf Time per solution} reflects the average time to discover an individual solution, as captured in the first column of
Fig.\ref{fig:Heuristic_time_per_solution}, considering 1000 problem instances. We observe that
the parallel discovery heuristic consistently reduces the average time required to obtain individual solutions. By allowing variables with low constraint degrees to remain multi-valued, our solver can discover multiple solutions in a single run. As problem size increases, this effect becomes even more pronounced: larger, less tightly constrained graphs present greater opportunity for multi-solution states, and the heuristic results in a larger speedup on a per-solution basis, as the second column of Fig.\ref{fig:Heuristic_time_per_solution} reveals. Even for the smallest problem size considered (9 nodes), this speed-up remains above two orders of magnitude.

\noindent {\bf Total time to solution:} 
Our parallel discovery heuristic relaxes local constraints, which typically makes the energy landscape smoother. As a result, convergence to the first solution can take longer. \textit{Total time to solution} reflects the overall runtime to reach at least one valid solution, capturing this effect. As shown in Fig.\ref{fig:Heuristic_total_time}, the slowdown, i.e., the increase in total time to solution is typically less than an order of magnitude, while the per-solution speedup (Fig.\ref{fig:Heuristic_time_per_solution}) exceeds several orders of magnitude. This highlights our heuristic's key trade-off: slower convergence to a single solution, but much faster discovery of multiple solutions.

\noindent {\bf Overhead analysis:}  
We next quantify the overhead induced by our heuristic
when it is unable to exploit problem structure. 
For most cases in our benchmark suite, the heuristic produced multiple solutions as expected. However, we observed two atypical patterns.

First, in a small fraction of instances ($\sim$1.25\%), the heuristic occasionally returned only a single solution, despite other trials on the same instances producing multiples. The explanation is in the probabilistic nature of our heuristic: multi-valued assignments are encouraged but not guaranteed, so rare singleton outcomes are expected and not problematic.  

Second, in an even smaller fraction ($\sim$0.4\%), the heuristic consistently produced only one solution. 
These instances were not single-solution CSPs -- verification with a complete CSP solver confirmed that multiple solutions existed. In this case the issue was structural: these graphs often contained cycles where our degree-based heuristic could not exploit the available flexibility.

\begin{figure*}[htbp]
    \centering
    \includegraphics[width=.91\linewidth]{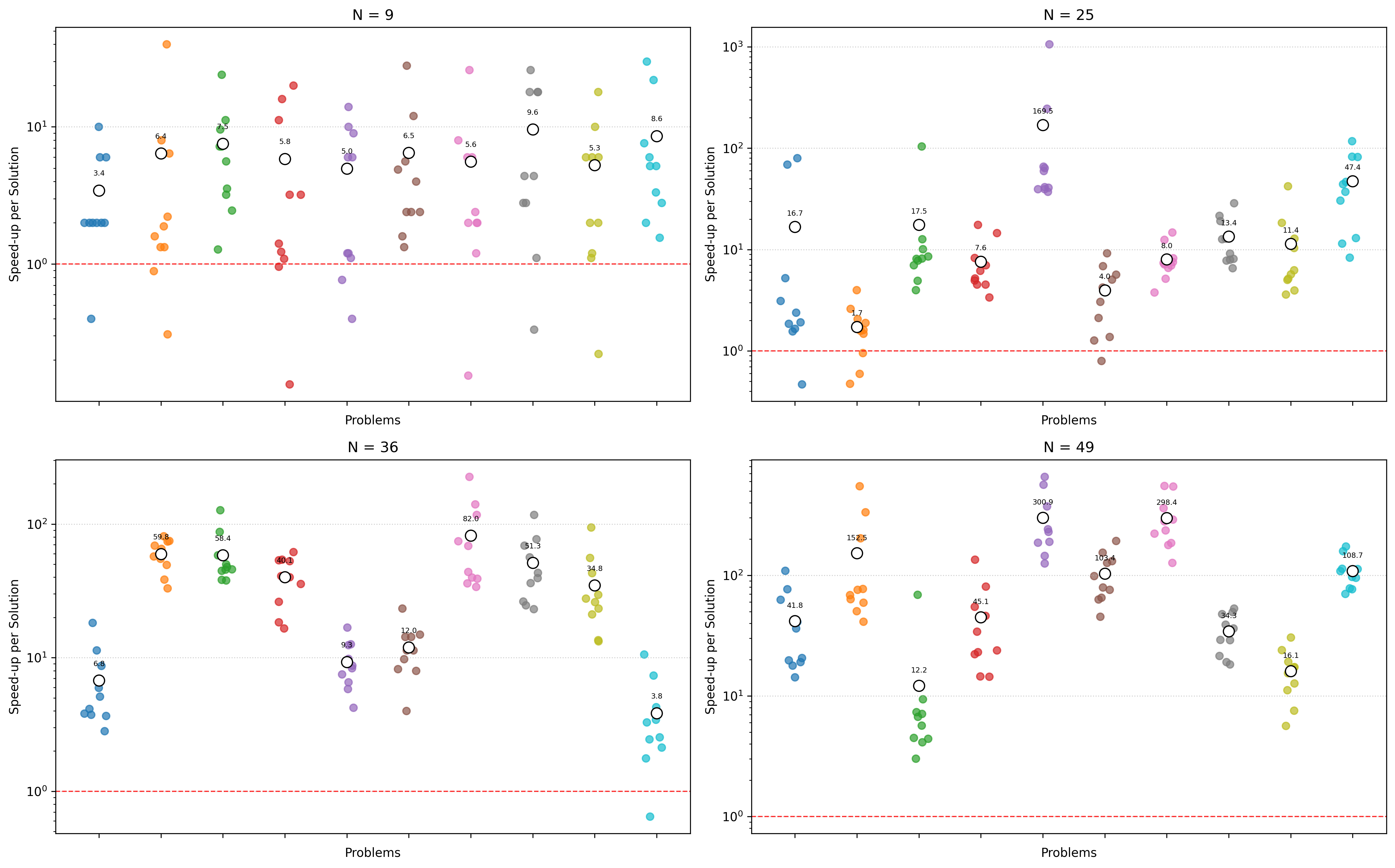}
    \caption{Comparison of our heuristic vs. sequential baseline runs for multi-solution discovery. Leveraging native parallelism in neuromorphic computing, our heuristic is consistently faster.}
    \label{fig:GC_Heuristic_Overhead}
\end{figure*}

To test genuinely single-solution cases, we generated problem instances verified with a complete solver to admit exactly one valid coloring. Such instances are extremely rare in 4-color planar graphs, but become more common under 3-color constraints. In these cases our solver naturally collapses to the baseline solver without the parallel discovery heuristic: when no degree slack exists, the WTA weights saturate and all variables are forced to singletons. 

In summary, we observed that 
(1) incidental singleton outcomes are rare and consistent with probabilistic sampling, (2) structural ``blind spots'' such as cycles simply reduce the heuristic to baseline behavior, and (3) in true single-solution problems the heuristic automatically degenerates to the non-heuristic solver. In all cases, no systematic slowdown was observed, confirming that the heuristic is robust even when inapplicable.

\noindent {\bf Parallel vs. sequential efficiency:} 
Finally, we compare our heuristic’s one-shot parallel multi-solution discovery against sequentially running the baseline solver (without the heuristic) multiple times to obtain the same number of solutions. Fig.\ref{fig:GC_Heuristic_Overhead} reports the speed-up of our heuristic solver over the baseline solver, and demonstrates that our heuristic is substantially faster,
often by several factors, since it exploits neuromorphic parallelism directly rather than relying on repeated independent runs. 

We also explore whether the solutions obtained are genuinely distinct. In graph-coloring, all valid solutions are of equal quality. Therefore,  they may differ only by color permutation (isomorphic solutions) or, in the worst case, they may be identical. In analyzing the diversity of solutions we particularly focus on smaller instances where the total solution count is limited. In 4-color problems up to 9 nodes, sequential reruns sometimes returned solutions from the same isomorphic group, and occasionally even exact duplicates. In contrast, our heuristic by construction cannot return multiple solutions from the same isomorphism class: the multi-valued representation forces exploration of distinct groups and rules out duplicates, even across isomorphic classes. For larger graphs, where the solution space grows rapidly, duplicates were not observed for either method.  

This demonstrates that our heuristic not only accelerates multi-solution discovery but also guarantees structural diversity of the returned solutions, reinforcing its advantage over sequential reruns.

\subsection{Comparison to Neuromorphic CSP Solvers}

\noindent None of the existing neuromorphic solvers from the literature systematically leverages parallel solution discovery in stark contrast to our solver. Therefore, by definition, such neuromorphic alternatives are only comparable to our baseline solver without the parallel discovery heuristic. We should note that, in principle, any neuromorphic solver can adopt our proposed parallel discovery heuristic.

In the following, we will cover how our baseline solver \textit{without the heuristic} compares to the most representative neuromorphic alternatives. For this analysis we use the exact same standard benchmark problems as the respective alternatives, as detailed in Section~\ref{Evaluation Setup}. 
For SAT, our time-to-solution scales exponentially with problem size as reported by Jonke et al.~\cite{jonke_solving_2016}. In graph coloring, our baseline solver converges much faster than SpiNNaker ($\approx17\times$ on Canada, $193\times$ on Australia, and $112\times$ on World instances)
~\cite{fonseca_guerra_using_2017}, and it also outperforms the standard WTA network of the Neo-Cortical approach ($\approx2\times$, $11\times$, $18\times$, and $24\times$ faster for graph sizes 9, 25, 36, and 49, respectively)~\cite{rutishauser_solving_2018}. For Sudoku, our baseline solver consistently outperforms SpiNNaker in convergence time ($\approx2.6\times$ on easy, $6.1\times$ on hard, and $8.2\times$ on AI Escargot instances), and for Ising spin benchmarks, it achieves lower variance and up to 3 orders of magnitude faster time-to-solution compared to SpiNNaker.

%% file: Texts/Related_Work.tex
\noindent The most directly related line of research to our study relies on spiking neural networks
for stochastic local search. 
Early work demonstrated that networks of stochastically spiking neurons could perform probabilistic inference, with Habenschuss et al.~\cite{habenschuss_stochastic_2013} and Jonke et al.~\cite{jonke_solving_2016} showing proof-of-concept applications of neural sampling to CSPs such as Traveling Salesman and Boolean Satisfiability. 
Subsequent efforts extended these ideas to hardware: Binas et al.~\cite{binas_spiking_2016} exploited thermal noise in analog VLSI neurons, while Fonseca Guerra et al.~\cite{fonseca_guerra_using_2017} demonstrated a SpiNNaker-compatible software framework. 
{Further, Davies et al.~\cite{Davies_Advancing_2021} realized neural sampling on Intel's Loihi neuromorphic processor and reported higher energy efficiency and lower latency compared to classical algorithms run on a CPU.} Rutishauser et al.~\cite{rutishauser_solving_2018} provided a theoretical analysis of WTA circuits, proving stability and convergence properties under different inhibitory schemes. 
While their networks are deterministic rather than sampling-based, their results set
the foundation for using WTA motifs as CSP constraints and for scaling such neuromorphic architectures.
These works collectively highlight the feasibility of mapping CSPs to neuromorphic substrates, but none exploit structural heuristics to accelerate solution discovery, which is the focus of our paper.
Beyond sampling-based formulations, other neuromorphic solvers have pursued algorithmic mappings. 
For example, Yakopcic et al.~\cite{yakopcic_leveraging_2020} demonstrated a quasi-complete CSP solver on Intel’s Loihi 2 by directly implementing the WalkSAT algorithm. 
Their approach exploits Loihi’s manycore architecture by instantiating multiple solver networks in parallel, effectively replicating WalkSAT across cores. 
This differs fundamentally from our method, which leverages the intrinsic dynamics of spiking networks to explore multiple solutions within a single network, rather than through replication.

%% file: Texts/Conclusion.tex
\noindent 
In this paper we introduce a parallel discovery heuristic 
for neuromorphic CSP solvers. We quantitatively demonstrate the effectiveness of the proposed heuristic in leveraging the native parallelism in neuromorphic representations to discover multiple solutions in a single run.  
Our evaluation on standardized CSP benchmarks shows that the heuristic can unlock more than  two orders-of-magnitude speed-up in 
finding a solution on average.
Importantly, the heuristic does not incur any significant overhead when problem structure prevents multi-solution states. As a result, the heuristic consistently outperforms repeated sequential runs to achieve the same, while preserving the structural diversity of solutions.

%% file: references.bib
@article{buesing_neural_2011,
	title = {Neural {Dynamics} as {Sampling}: {A} {Model} for {Stochastic} {Computation} in {Recurrent} {Networks} of {Spiking} {Neurons}},
	volume = {7},
	issn = {1553-7358},
	shorttitle = {Neural {Dynamics} as {Sampling}},
	url = {https://dx.plos.org/10.1371/journal.pcbi.1002211},
	doi = {10.1371/journal.pcbi.1002211},
	language = {en},
	number = {11},
	urldate = {2024-12-27},
	journal = {PLoS Computational Biology},
	author = {Buesing, Lars and Bill, Johannes and Nessler, Bernhard and Maass, Wolfgang},
	editor = {Sporns, Olaf},
	month = nov,
	year = {2011},
	pages = {e1002211},
}

@article{jonke_solving_2016,
	title = {Solving {Constraint} {Satisfaction} {Problems} with {Networks} of {Spiking} {Neurons}},
	volume = {10},
	issn = {1662-453X},
	doi = {10.3389/fnins.2016.00118},
	language = {English},
	urldate = {2024-12-11},
	journal = {Frontiers in Neuroscience},
	author = {Jonke, Zeno and Habenschuss, Stefan and Maass, Wolfgang},
	month = mar,
	year = {2016},
	note = {Publisher: Frontiers},
}

@inproceedings{binas_spiking_2016,
	title = {Spiking analog {VLSI} neuron assemblies as constraint satisfaction problem solvers},
	url = {https://ieeexplore.ieee.org/document/7538992},
	doi = {10.1109/ISCAS.2016.7538992},
	urldate = {2025-05-13},
	booktitle = {2016 {IEEE} {International} {Symposium} on {Circuits} and {Systems} ({ISCAS})},
	author = {Binas, Jonathan and Indiveri, Giacomo and Pfeiffer, Michael},
	month = may,
	year = {2016},
	note = {ISSN: 2379-447X},
	pages = {2094--2097},
}

@article{fonseca_guerra_using_2017,
	title = {Using {Stochastic} {Spiking} {Neural} {Networks} on {SpiNNaker} to {Solve} {Constraint} {Satisfaction} {Problems}},
	volume = {11},
	issn = {1662-4548},
	doi = {10.3389/fnins.2017.00714},
	language = {eng},
	journal = {Frontiers in Neuroscience},
	author = {Fonseca Guerra, Gabriel A. and Furber, Steve B.},
	year = {2017},
	pmid = {29311791},
	pmcid = {PMC5742150},
	pages = {714},
}

@article{rutishauser_solving_2018,
	title = {Solving {Constraint}-{Satisfaction} {Problems} with {Distributed} {Neocortical}-{Like} {Neuronal} {Networks}},
	volume = {30},
	issn = {1530-888X},
	doi = {10.1162/NECO_a_01074},
	language = {eng},
	number = {5},
	journal = {Neural Computation},
	author = {Rutishauser, Ueli and Slotine, Jean-Jacques and Douglas, Rodney J.},
	month = may,
	year = {2018},
	pmid = {29566357},
	pmcid = {PMC5930080},
	pages = {1359--1393},
}

@inproceedings{yakopcic_leveraging_2020,
	address = {Glasgow, United Kingdom},
	title = {Leveraging the {Manycore} {Architecture} of the {Loihi} {Spiking} {Processor} to {Perform} {Quasi}-{Complete} {Constraint} {Satisfaction}},
	copyright = {https://ieeexplore.ieee.org/Xplorehelp/downloads/license-information/IEEE.html},
	isbn = {978-1-72816-926-2},
	url = {https://ieeexplore.ieee.org/document/9207419/},
	doi = {10.1109/IJCNN48605.2020.9207419},
	language = {en},
	urldate = {2024-12-12},
	booktitle = {2020 {International} {Joint} {Conference} on {Neural} {Networks} ({IJCNN})},
	publisher = {IEEE},
	author = {Yakopcic, Chris and Rahman, Nayim and Atahary, Tanvir and Taha, Tarek M. and Douglass, Scott},
	month = jul,
	year = {2020},
	pages = {1--8},
}

@article{habenschuss_stochastic_2013,
	title = {Stochastic {Computations} in {Cortical} {Microcircuit} {Models}},
	volume = {9},
	issn = {1553-7358},
	url = {https://dx.plos.org/10.1371/journal.pcbi.1003311},
	doi = {10.1371/journal.pcbi.1003311},
	language = {en},
	number = {11},
	urldate = {2024-12-27},
	journal = {PLoS Computational Biology},
	author = {Habenschuss, Stefan and Jonke, Zeno and Maass, Wolfgang},
	editor = {Sporns, Olaf},
	month = nov,
	year = {2013},
	pages = {e1003311},
}

@ARTICLE{Pedretti_phase_change_2020,
  author={Pedretti, Giacomo and Mannocci, Piergiulio and Hashemkhani, Shahin and Milo, Valerio and Melnic, Octavian and Chicca, Elisabetta and Ielmini, Daniele},
  journal={IEEE Journal on Exploratory Solid-State Computational Devices and Circuits}, 
  title={A Spiking Recurrent Neural Network With Phase-Change Memory Neurons and Synapses for the Accelerated Solution of Constraint Satisfaction Problems}, 
  year={2020},
  volume={6},
  number={1},
  pages={89-97},
  doi={10.1109/JXCDC.2020.2992691}}

@article{Aimone_non_cognitive_NC_2022,
title = "A review of non-cognitive applications for neuromorphic computing",
author = "Aimone, {James B.} and Prasanna Date and Fonseca-Guerra, {Gabriel A.} and Hamilton, {Kathleen E.} and Kyle Henke and Bill Kay and Kenyon, {Garrett T.} and Kulkarni, {Shruti R.} and Mniszewski, {Susan M.} and Maryam Parsa and Risbud, {Sumedh R.} and Schuman, {Catherine D.} and William Severa and Smith, {J. Darby}",
note = "Publisher Copyright: {\textcopyright} 2022 The Author(s). Published by IOP Publishing Ltd.",
year = "2022",
month = sep,
day = "1",
doi = "10.1088/2634-4386/ac889c",
language = "English",
volume = "2",
journal = "Neuromorphic Computing and Engineering",
issn = "2634-4386",
publisher = "IOP Publishing",
number = "3",
}

@ARTICLE{Davies_Advancing_2021,
  author={Davies, Mike and Wild, Andreas and Orchard, Garrick and Sandamirskaya, Yulia and Guerra, Gabriel A. Fonseca and Joshi, Prasad and Plank, Philipp and Risbud, Sumedh R.},
  journal={Proceedings of the IEEE}, 
  title={Advancing Neuromorphic Computing With Loihi: A Survey of Results and Outlook}, 
  year={2021},
  volume={109},
  number={5},
  pages={911-934},
  doi={10.1109/JPROC.2021.3067593}}

@inproceedings{Hamdioui_Data_Intensive_2015,
author = {Hamdioui, Said and Xie, Lei and Nguyen, Hoang Anh Du and Taouil, Mottaqiallah and Bertels, Koen and Corporaal, Henk and Jiao, Hailong and Catthoor, Francky and Wouters, Dirk and Eike, Linn and van Lunteren, Jan},
title = {Memristor based computation-in-memory architecture for data-intensive applications},
year = {2015},
isbn = {9783981537048},
publisher = {EDA Consortium},
address = {San Jose, CA, USA},
booktitle = {Proceedings of the 2015 Design, Automation \& Test in Europe Conference \& Exhibition},
pages = {1718–1725},
numpages = {8},
location = {Grenoble, France},
series = {DATE '15}
}

@article{Boltzmann_1989,
title = {Combinatorial optimization on a Boltzmann machine},
journal = {Journal of Parallel and Distributed Computing},
volume = {6},
number = {2},
pages = {331-357},
year = {1989},
issn = {0743-7315},
doi = {https://doi.org/10.1016/0743-7315(89)90064-6},
url = {https://www.sciencedirect.com/science/article/pii/0743731589900646},
author = {Jan H.M. Korst and Emile H.L. Aarts}
}

@book{tsang_foundations_1993,
	series = {Computation in cognitive science},
	title = {Foundations of {Constraint} {Satisfaction}},
	isbn = {978-0-12-701610-8},
	url = {https://books.google.com/books?id=TnxQAAAAMAAJ},
	publisher = {Academic Press},
	author = {Tsang, E.},
	year = {1993},
	lccn = {gb93066472},
}

@article{Elliot_Tree_Search_1980,
author = {Elliot, G.},
year = {1980},
month = {01},
pages = {},
title = {Increasing tree search efficiency for constraint satisfaction},
volume = {14},
journal = {Artificial Intelligence - AI}
}

@ARTICLE{Marques_Clause_Conflict_1999,
  author={Marques-Silva, J.P. and Sakallah, K.A.},
  journal={IEEE Transactions on Computers}, 
  title={GRASP: a search algorithm for propositional satisfiability}, 
  year={1999},
  volume={48},
  number={5},
  pages={506-521},
  doi={10.1109/12.769433}}

@article{Mackworth_Consistency_1977,
title = {Consistency in networks of relations},
journal = {Artificial Intelligence},
volume = {8},
number = {1},
pages = {99-118},
year = {1977},
issn = {0004-3702},
doi = {https://doi.org/10.1016/0004-3702(77)90007-8},
url = {https://www.sciencedirect.com/science/article/pii/0004370277900078},
author = {Alan K. Mackworth}
}

@InProceedings{Sabin_Constraint_Propagation_Backtracking_1994,
author="Sabin, Daniel
and Freuder, Eugene C.",
editor="Borning, Alan",
title="Contradicting conventional wisdom in constraint satisfaction",
booktitle="Principles and Practice of Constraint Programming",
year="1994",
publisher="Springer Berlin Heidelberg",
address="Berlin, Heidelberg",
pages="10--20",
isbn="978-3-540-49032-6"
}

@inproceedings{Selman_GSAT_1992,
author = {Selman, Bart and Levesque, Hector and Mitchell, David},
title = {A new method for solving hard satisfiability problems},
year = {1992},
isbn = {0262510634},
publisher = {AAAI Press},
booktitle = {Proceedings of the Tenth National Conference on Artificial Intelligence},
pages = {440–446},
numpages = {7},
location = {San Jose, California},
series = {AAAI'92}
}

@inproceedings{selman_noise_1994,
	address = {Seattle, Washington},
	series = {{AAAI}'94},
	title = {Noise strategies for improving local search},
	urldate = {2025-09-07},
	booktitle = {Proceedings of the {Twelfth} {AAAI} {National} {Conference} on {Artificial} {Intelligence}},
	publisher = {AAAI Press},
	author = {Selman, Bart and Kautz, Henry A. and Cohen, Brain},
	month = aug,
	year = {1994},
	pages = {337--343},
}

@inproceedings{hoos_satlib_2000,
  author    = {Holger H. Hoos and Thomas St{\"u}tzle},
  title     = {SATLIB: An Online Resource for Research on SAT},
  booktitle = {SAT 2000},
  editor    = {I. P. Gent and H. van Maaren and T. Walsh},
  pages     = {283--292},
  publisher = {IOS Press},
  year      = {2000},
  url       = {http://www.satlib.org}
}
